# Optimizing supercontinuum spectro-temporal properties by leveraging machine learning towards multi-photon microscopy

**Van Thuy Hoang, Yassin Boussafa, Lynn Sader, Sébastien Février, Vincent Couderc, Benjamin Wetzel***

XLIM Research Institute, CNRS UMR 7252, Université de Limoges, Limoges 87060, France

**Correspondence:**
Corresponding Author
benjamin.wetzel@xlim.fr

**Abstract**

Multi-photon microscopy has played a significant role in biological imaging since it allows to observe living tissues with improved penetration depth and excellent sectioning effect. Multi-photon microscopy relies on multi-photon absorption, enabling the use of different imaging modalities that strongly depends on the properties of the sample structure, the selected fluorophore and the excitation laser. However, versatile and tunable laser excitation for multi-photon absorption is still a challenge, limited by e.g. the narrow bandwidth of typical laser gain medium or by the tunability of wavelength conversion offered by optical parametric oscillators or amplifiers. As an alternative, supercontinuum generation can provide broadband excitationspanning from the ultra-violet to far infrared domains and integrating numerous fluorophore absorption peaks, in turn enabling different imaging modalities or potential multiplexed spectroscopy. Here, we report on the use of machine learning to optimize the spectro-temporal properties of supercontinuum generation in order to selectively enhance multi-photon excitation signals compatible with a variety of fluorophores (or modalities) for multi-photon microscopy. Specifically, we numerically explore how the use of reconfigurable (femtosecond) pulse patterns can be readily exploited to control the nonlinear propagation dynamics and associated spectral broadening occurring in a highly-nonlinear fiber. In this framework, we show that the use of multiple pulses to seed optical fiber propagation can trigger a variety of nonlinear interactions and complex propagation scenario. This approach, exploiting the temporal dimension as an extended degree of freedom, is used to maximize typical multi-photon excitations at selected wavelengths, here obtained in a versatile and reconfigurable manner suitable for imaging applications. We expect these results to pave the way towards on-demand and real time supercontinuum shaping, with further multi-photon microscopy improvements in terms of spatial 3D resolution, optical toxicity, and wavelength selectivity.

# 1    Introduction

Multi-photon excitation (MPE) microscopy has become essential tool for studying biological matter as it allows selective imaging, considering the 4D (x-y-z-t) examination of biological living tissues (Zipfel et al., 2003). Unlike one-photon excitation fluorescence microscopy techniques where the excitation wavelengths typically span the ultraviolet and visible region (Cheng, et al. 2002; Campagnola et al., 2003; Rodriguez et al., 2006; Evans et al., 2008), MPE microscopy is implemented using a longer wavelength range (i.e., visible and near-infrared) resulting in a reduction of scattering loss and photodamage of the tissue sample, while providing excellent sectioning effect to ensure sufficient spatial resolution. MPE microscopy has therefore been used to image living tissues with further penetration depth (Helmchen et al., 2005; Larson, 2011) and to accurately probe the quantitative structure and function of cellular activities with sub-micro resolution (Zhao et al. 2014; Borhani et al. 2019). Recently, thanks to the advance in high-power femtosecond lasers, MPE microscopy has for instance been successfully reported for *in-vivo* deep brain imaging with a penetration depth above the mm range (Kobat et al. 2011; Horton et al., 2013; Wang et al., 2020; Weisenburger et al., 2019; Wu et al., 2021).

Beyond multi-photon absorption (MPA), MPE microscopy also relies on nonlinear parametric processes, which involve both multiphoton absorption and nonlinear wave mixing. These phenomena strongly depend on the sample structural constitution, the laser beam characteristics (e.g., pulse peak power, wavelength, spatio-temporal coherence), but also on the properties of the selected fluorophores (e.g., multi-photon excitation cross-section) when using labelled samples (Xu et al., 1996; Larson, 2011; Lefort, 2017). For instance, a laser source with high pulse energy provides a high photon flux, so that to mitigate scattering loss and thus increase the reachable imaging depth. However, such high energy pulses will inadvertently increase tissue heating, therefore resulting in damaging effects of the living tissues and their associated biological processes. As a consequence, adapting the laser illumination properties to match the selected fluorophore excitation requirements (or any other related multi-photon imaging modality such as second- or third-harmonic generation (Sheppard 2020) is a necessary step to improve the penetration depth and resolution of MPE microscopy. Nevertheless, in reality, having a versatile and tunable laser to meet these versatile specifications is a challenge, limited by e.g., the narrow bandwidth of the gain medium used in fiber or bulk laser systems.

As a consequence, only a few of the vastly available excitable fluorophores whose MPA peaks match fixed laser wavelength emissions can be readily used for MPE microscopy. To circumvent this issue, other widespread approaches consist in using processes of harmonic generation and optical parametric oscillators (OPO) or amplifiers (OPA) to adjust the excitation wavelengths required for multi-photon imaging techniques (Chu et al., 2003; Kobat et al., 2011). The drawbacks of these methods are the inherently complex and expensive setups used for practical experimental implementations, usually paired with several difficulties to operate and the requirement for a special maintenance.

As an alternative, supercontinuum (SC) generation has been implemented for MPE microscopy. Indeed, SC sources can provide a broadband wavelength spectrum (i.e., from ultraviolet to mid-IR) covering numerous MPA peaks of the fluorophores (Poudel et al., 2019) or being directly relevant for multiplexed spectroscopy (Labruyère et al., 2012; Falconieri et al., 2019).

For instance, Coherent Anti-Stokes Raman Scattering (CARS), which was first studied in the sixties (Maker et al., 1965; Begley et al., 1974), is a four-wave mixing phenomenon based on the identification of molecular vibrational modes that allows the structural cartography of labelled-free biological samples. Later on, that concept has been developed both in collinear and non-collinear geometry for vibrational signature identification in the CH region and for studying organic liquids, respectively (Zumbuch et al., 1999; Duncan et al., 1982).

Multiplex-CARS (M-CARS) experiments, based on simultaneous excitation of all frequencies ranging from 0 to 4500 cm$^{-1}$ (using a monochromatic pump beam and a large band Stokes wave), have been demonstrated in 2002, by mixing polychromatic and monochromatic laser sources (Müller et al., 2002). In this framework, supercontinuum sources appeared as the best way to implement M-CARS systems (Okuno et al., 2007; Klarskov et al., 2011; Shen et al., 2018). However, because of the group velocity dispersion encountered in optical fibers, laser sources with long pulses have been priviledged to guaranty temporal overlap between all the frequencies constituting the continuum and the quasi-monochromatic pump signals (Okuno et al., 2007). Yet, broadband SC has been scarcely implemented (directly) for multi-photon imaging because the residual part of the SC spectrum may give rise to additional noise within the MPA signals, on top of bringing undesired photodamage to the sample.

In fact, almost all MPA applications, including M-CARS techniques, rely on extracting a restricted range or only a few of the desired excitation wavelengths from the SC spectrum by using optical filtering techniques (Poudel et al., 2019). In such a case, however, only a small portion of the SC is extracted and used to generate the fluorescence signal, thus discarding most of the SC energy. More importantly, in numerous SC sources, the requirement of generating an extremely broadband signal with sufficient power spectral density over the whole SC spectral coverage is associated with dramatic drawbacks in terms of power efficiency. Indeed, in most commercial sources, these SC are generated from relatively long pulses (typically nanosecond duration) to bring about sufficient energy for extreme spectral broadening. Interestingly, however, MPE fluorescence signals are intrinsically related to the peak power of the excitation thus making the use of long pulses not always ideal for multiphoton microscopy. In fact, while the ultimate optical excitation source may be up for debate, it is clear that a versatile and reconfigurable system is highly sought-after. In this framework, customizing the properties of SC generation to maximize the power and the temporal waveform shape at suitably selected wavelengths thus constitute an essential requirement.

As a succinct reminder, it is worth noting that SC generation can be seen as the result of dispersive and nonlinear effects acting together in complex fashion during pulse propagation, typically in a guided structure such as a fiber or an integrated waveguide (Dudley and Taylor, 2010). As such, it is thus possible to tune SC spectro-temporal properties by changing e.g. the fiber structure, the fiber material, or the parameters of the input laser pulses. The first two approaches, however, do not exhibit proper methods for "on the fly" control properties of SC generation without significantly modifying the setup. In contrast, controlling nonlinear propagation by changing the input pulse laser parameters (i.e., pulse duration, peak power, chirp) has been used as a convenient way to optimize SC generation for practical applications, in particular, due to the strong dependence and sensitivity of spectral broadening dynamics on the input laser source parameters (Veljković et al., 2019; Sylvestre et al., 2021). Interestingly, such input laser parameters may be automatically modified via active systems (such as amplifiers, programmable filters, modulators, polarization controllers) enabling the implementation of optimization algorithm and thus allowing for live and autonomous adjustment of the desired SC properties. Recently, there has been a growing interest in applying various machine learning strategies to predict and control propagation dynamics in multidimensional nonlinear systems (Genty et al., 2021). For instance, deep neural networks have been deployed to predict light transmission properties of multimode fibers for imaging techniques (Kakkava et al., 2019), or to reconstruct a distorted image from the intensity of the output speckle patterns in such fibers (Borhani et al., 2018). For nonlinear propagation, a feed-forward neural network was also used to accurately predict the temporal profile of red-shifted solitons (Salmela et al., 2020) and the more global nonlinear propagation dynamics occurring during SC generation (Salmela et al., 2021).

For the adjustment of optical pulse properties, multiple strategies can be envisioned, including widespread pulse shaping techniques, chirp and dispersion engineering, etc. (Weiner 2011) Among those, an innovative approach based on integrated photonics has been demonstrated over the last years, by using cascaded on-chip Mach-Zehnder interferometers (MZIs) in combination with machine learning algorithms. This strategy was for instance instrumental for the demonstration of picosecond waveform laser pulses being autonomously reconfigured (Fischer et al., 2021). In a similar framework, a genetic algorithm (GA) optimization was recently used for the formation of reconfigurable pulse patterns so that to maximize the intensity of selected SC wavelengths after nonlinear fiber propagation (Wetzel et al., 2018). Such pulse patterns were created via a femtosecond pulse and an integrated photonic chip made of cascaded MZIs: the suitable modification of MZI splitting ratios allows for the generation of temporally interleaved pulse replica and, consequently, the control of SC spectral properties after nonlinear broadening. Unlike single-pulse seeds, the use of pulse patterns provides additional degrees of freedom (similar to a multidimensional system) to optimize the output properties of the SC generation via a variety of nonlinear phenomena including both intra-pulse effects and complex inter-pulse interactions (Andresen et al., 2011, Wetzel et al., 2018).

However, from an experimental perspective, several key properties of MPE microscopy (such as multi-photon excitation lifetime and cross-section, optical toxicity, etc.) not only rely on the peak power, but also on the temporal waveform of the excitation pulses. In addition, the requirement for advanced temporal waveform control is further highlighted by the rise/fall time of the detectors, which is typically in the order of hundreds of picoseconds. For an improved applicability in MPE microscopy, there is therefore an urgent need for novel yet easy and scalable approaches alowing for the fine tuning and optimization of SC spectro-temporal properties beyond simple filtering and/or power spectral density enhancement.

In this article, we conduct a numerical study to assess the viability of temporal pulse pattern shaping for MPE microscopy. Specifically, we implement various machine learning strategies to maximize MPA signals and waveform properties at specific wavelengths in the SC spectrum. In particular, we show how the use of ultrafast pulse patterns generated from a cascaded MZIs chip can be leveraged to adjust the output temporal waveform for a variety of MPA processes. The on-chip MZIs are used as an analogous to a multidimensional pulse pattern processor to shape the temporal waveforms (before nonlinear propagation). In turn, this allows for tuning the complex nonlinear dynamics of pulse broadening toward applications in MPE microscopy. We point out that, considering SC generation in a highly nonlinear fiber (HNLF), this strategy gives rise to improved flexibility compared to single pulse seeds with reconfigurable properties: we demonstrate the shaping of output pulses with a variety of temporal waveforms for two-photon absorption (2PA), three-photon absorption (3PA), and conjoint 2PA-3PA. Taking advantage of machine learning, the desired MPA signals can be maximized, with the temporal profiles and coherence degree of the waveform suitably adjusted.

## 2 Material and Methods

### 2.1 Numerical modeling of nonlinear pulse propagation

For our numerical study, we use a split-step Fourier method to solve the general nonlinear Schrödinger equation (GNLSE), as given in **Eq. (1)**:

$$\frac{\partial A}{\partial z} + \frac{\alpha}{2} A - \sum_{k \geq 2} \frac{i^{k+1}}{k!} \beta_k \frac{\partial^k A}{\partial T^k} = i\gamma \left(1 + \tau_{shock} \frac{\partial}{\partial T}\right) \left(A(z,T) \int_{-\infty}^{\infty} R(T') |A(z, T-T')|^2 dT'\right) \quad (1)$$

where $A(z,T)$ is the pulse envelope and $T = t - \beta_1 z$ is the time coordinate in the comoving frame at the group velocity $\beta_1^{-1}$. The right-hand side of **Eq. (1)** describes nonlinear effects including Kerr and Raman effects in which γ is the nonlinear coefficient, the shock timescale $\tau_{shock} = 1/\omega_0$ is to present the self-steepening. The nonlinear response function $R(T') = (1 - f_R)\delta(T') + f_R h_R(T')$ includes the instantaneous Kerr effect and delayed Raman contribution in which $f_R$ = 0.18 accounts for the Raman fraction. The left-hand side of **Eq. (1)** describes linear effects including loss (α) and dispersion ($\beta_k$).

Note that this GNLSE approach is considered for numerically simulating pulse evolution in both a highly nonlinear fiber (HNLF) and the waveguides constituting our on-chip programmable delay line (PDL - see details below).

The vacuum noise (shot noise) is taken into account in our model in **Eq. (1)** by adding one photon with a random phase on each spectral bin of the input waveform (Dudley et al., 2006). Herein, we do not consider additional noise effects, such as the amplitude fluctuation of each laser pulse (i.e. laser technical noise), polarization noise, etc., which typically degrade the coherence of the output SC spectrum. However, we assume that the impact of such noise contributions is limited when considering short laser pulse duration in the femtosecond regime (Zhu and Brown, 2004; Genier et al., 2019). The coherence degree $|g_{12}^1(\lambda, 0)|$ of the filtered spectrum at the wavelengths selected for MPA is calculated with a fixed value of splitting ratio (so that to generate the same pulse pattern) but with 20 simulations with a random noise (i.e. vacuum noise with one photon with random phase per spectral bin). The average coherence $< |g_{12}^1(\lambda, 0)| >$ is considered as the mean value of coherence integrated over the bandwidth of the filtered spectrum.

For spectral broadening, we consider the nonlinear propagation of optical pulses into 10 m of HNLF. The parameters are those retrieved from a home-made (single-mode) Ge-doped fiber available in our laboratory, similar to typical commercial HNLF with a zero dispersion in the C-band around 1550 nm. Considering a central wavelength at 1560 nm, the fiber has a nonlinear coefficient of 3.5 W$^{-1}$km$^{-1}$ and dispersion is located in the anomalous regime with $\beta_2$ = -2.15 ps$^2$/km and $\beta_3$ = 0.0693 ps$^3$/km. The fiber losses at wavelengths of 1200 nm, 1560 nm and 2000 nm are taken as 60 dB/km, 1 dB/km, and 100 dB/km, respectively.

For the initial processing of the input pulse prior to HNLF injection (so that to generate versatile temporal patterns of multiple femtosecond pulses), we also consider the propagation of the initial laser pulses into our on-chip programmable delay line: the considered PDL structure comprises 8 cascaded unbalanced interferometers (with increasing delays) to split an initial single pulse into up to 256 individual pulses with 1 ps separation between two adjacent ones, as reported in (Wetzel et al., 2018) and (Fischer et al., 2021). Each interferometer is an unbalanced interferometric structure (made of two integrated optical waveguides, two 50:50 optical couplers and a phase shifter) to constitute a balanced MZI followed by a pair of unbalanced waveguides. Each of these allows for splitting the incoming pulse(s) and delaying its (their) replicas by only changing the MZI splitting ratio (i.e., by adjusting the relative phase difference between the waveguide arms of the MZI).

A complete modelisation of the pulse propagation in the PDL is carried out by considering two GNLSE simulations (one for each waveguide) which are periodically coupled at the location of the MZI waveguide couplers while considering the phase shift implemented on the respective MZI arms. For this on-chip pulse splitting process on the PDL, we assume that the light propagating in each integrated waveguide is linearly polarized and follows a purely single-mode guiding condition (in a TM polarization mode). The PDL waveguides are therefore considered based on realistic experimental conditions so that at 1560 nm, the dispersion parameters are $\beta_2$ = -2.87 ps$^2$/km and $\beta_3$ = -0.0224 ps$^3$/km, the nonlinear coefficient is 233 W$^{-1}$km$^{-1}$ and the losses of 600 dB/km remain marginal for the propagation in the ~5 cm-long waveguide structure (Ferrera et al., 2008; Wetzel et al., 2018).

In our model, we consider ideal MZI response (with perfect extinction ratio) so that the splitting ratio of each interferometer can vary over the range 0 to 1, in which a value of 0 indicates that the whole incoming light is routed into the following "short arm" section of the unbalanced interferometer, and a value of 1 conversely indicates a routing into the following "long arm" of the interferometer. Changing the splitting ratio of each interferometer thus leads in shaping the waveform of the pulse patterns, as illustrated in **Figure 1**.

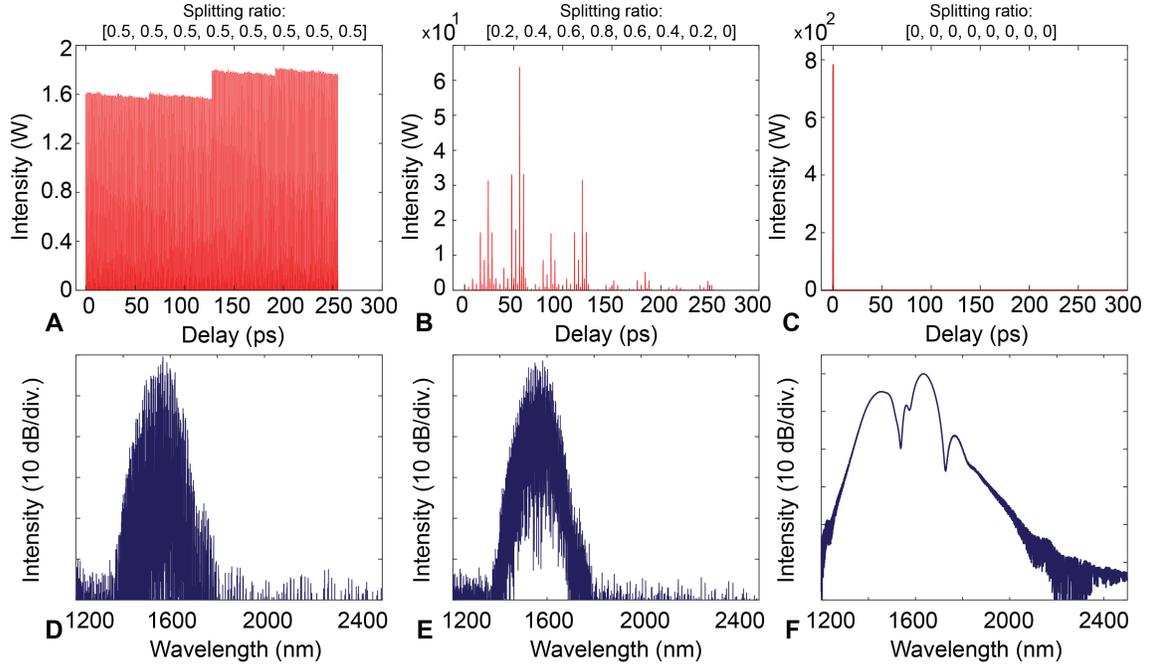

**FIGURE 1 |** Temporal (top) and spectral (bottom) profiles of the pulse patterns generated from the on-chip PDL for various splitting ratios, when initially injected with a 100 fs pulse at 1kW peak power: (A) and (D) show the equipartition of the input pulse energy into 256 pulses when the splitting ratio is 0.5 for each MZI; (B) and (E) show the generation of an arbitrary pulse pattern when the MZIs' splitting ratio are set to [0.2, 0.4, 0.6, 0.8, 0.6, 0.4, 0.2, 0]; (C) and (F) show the case where the whole input pulse energy remains in the first pulse 'slot' when the splitting ratio is 0 for each MZI (hence following the shortest path throughout propagation in the on-chip PDL).

In this study, we numerically consider two cases of nonlinear propagation into our HNLF, using either (i) single pulse seeds with adjustable parameters or (ii) reconfigurable pulse patterns as generated from our on-chip PDL. The initial laser pulses possess a Gaussian shape as shown in **Eq. (2)**:

$$A(0,T) = \sqrt{P_0} exp\left(-(1+iC)\frac{2ln(2)T^2}{T_0^2}\right) \quad (2)$$

where $P_0$ is the peak power, $T_0$ is the pulse duration (full width at half maximum - FWHM) and $C$ is the chirp factor.

For single pulse seeds, the spectro-temporal properties of SC generation are arbitrarily modified by adjusting the input pulse parameters: we consider a broad range of peak power spanning from 5 to 20 kW, a pulse duration over the range 50 – 200 fs, and a chirp factor adjustable from -10 to +10. We note that the pulse duration is here constrained to this 50 – 200 fs range to ensure a high degree of coherence of the output SC spectrum. The central wavelength of the laser source is 1560 nm.

For reconfigurable pulse patterns, we define a set of constant parameters for the initial pulse (i.e., 100 fs pulse duration, 1 kW peak power, 1560 nm central wavelength) injected into the on-chip PDL. The splitting ratio of each interferometer is modified to shape the temporal waveform of the pulse patterns at the PDL output. Subsequently, the generated temporal pulse patterns are amplified by 20 dB as typically expected from an erbium-doped fiber amplifier (EDFA) before injection into the HNLF for nonlinear propagation, see **Figure 2**.

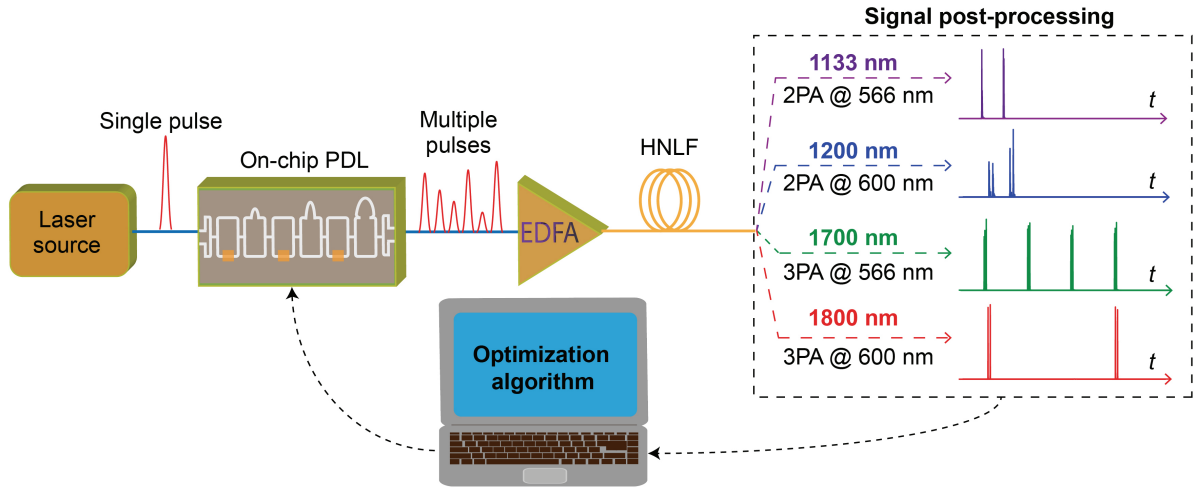

**FIGURE 2** | Schematic representation showing the use of reconfigurable temporal pulse patterns, as generated from an on-chip programmable delay line (PDL), to seed nonlinear propagation into a highly nonlinear fiber (HNLF) and spectral broadening. The interferometers settings on the PDL can be readily adjusted via a variety of machine learning approaches to optimize the output optical waveform and selectively enhance multi-photon absorption (MPA) signals at wavelengths suitable for e.g. bio-imaging applications (see the color code).

## 2.2 Optimization parameters and target functions for multiphoton excitation microscopy

As an illustrative example of our approach for the versatile optimization of multi-photon microscopy, we explore the potential application of reconfigurable SC for generating optical signals at wavelengths suitable for typical MPA excitations. For instance, we consider two sets of spectral regions respectively compatible with two- and three-photon absorption within the near-infrared region. In particular, we selectively filtered broadband signals with central wavelengths at (i) 1133 nm and 1700 nm, towards the excitation of 2PA and 3PA for fluorophores and endogenous fluorescent proteins (FPs) with a central wavelength of the fluorescence at 566 nm (e.g., Texas Red, mKate and tdKatushka2). We also considered the wavelengths at 1200 nm and 1800 nm for 2PA and 3PA processes with central wavelengths of fluorescence at 600 nm (e.g., Alexa Fluor 680), see color-coded wavelengths in **Figure 2**.

The selected fluorophores (Texas Red, Alexa Fluor) are extremely popular for several reasons: their low-cost, minimal aggregation issues, and low cytotoxicity. The FPs have been used to increase the brightness in biological settings as well as far-red emitters with substantial potential for deep *in-vivo* imaging (Miller et al., 2017; Ricard et al., 2018; Wang et al., 2018). Moreover, excitation pulses at these selected wavelengths ensure minimizing the losses of the MPA process, including both scattering loss and photon absorption by tissues (Miller et al., 2017). Such a low photon absorption enables a better penetration depth of multi-photon imaging with low input laser power and also mitigates deleterious thermal effects (i.e. tissue heating and photodamage) which can be extremely harmful to biological samples.

For simplicity, we here assume that the spectral bandwidth (FWHM) of the fluorescent spectra from both 2PA and 3PA for all selected fluorophores is 20 nm, and accordingly, the FWHM filtering bandwidth of the excitation pulses at 1133 and 1200 nm (i.e. 2PA excitations) is 40 nm, while the filtering bandwidth at 1700 and 1800 nm (i.e. 3PA excitations) is 60 nm. The excitation pulses are extracted from the SC spectrum (in the frequency domain) using a Gaussian filter function, and the resulting filtered waveforms are retrieved in the time-domain by a Fourier transform.

The target functions shown in **Eq. (3)** are used for the optimization process. They correspond to the filtered intensity squared for 2PA (cubed for 3PA) and integrated with respect to time

$$I_{2PA,\lambda_{1,2}} = \int (I_{\lambda_{1,2}})^2 dt \quad \text{and} \quad I_{3PA,\lambda_{3,4}} = \int (I_{\lambda_{3,4}})^3 dt \tag{3}$$

where $\lambda_{1,2,3,4}$ are 1133 nm, 1200 nm, 1700 nm, and 1800 nm, respectively. $I_{2PA,\lambda_{1,2}}$ and $I_{3PA,\lambda_{3,4}}$ correspond to the target functions related to the MPA signals for 2PA and 3PA, respectively.

The MPA signal according to the intensities in **Eq. (3)** is maximized by an optimization process based on evolutionary algorithms. In particular, we numerically implement two optimization techniques respectively based on a GA and particle swarm optimization (PSO), each possessing its own advantages (Fischer et al., 2021; Jiang, et al., 2010): GA optimization includes single-objective GA to maximize 2PA and 3PA signals, but also multi-objective GA functions for conjointly optimizing two individual MPA processes: 2PA-2PA or 3PA-3PA for two different fluorophores, as well as 2PA-3PA for a selected fluorophore. In the case of multi-objective GA, the optimization yields a Pareto front (also called Pareto frontier) that represents a trade-off between two MPA signal optimization.

In our GA model, the crossover fraction is set to 50%, and the population size to 1000 individuals (with 3 genes for single pulse seeds and 9 genes for reconfigurable pulse patterns, respectively) (Katoch et al., 2021). The number of generation for multi-objective GA is set to 30. For single-objective GA, the minimum of the fitness function is typically found within 15 generations as shown in the following section. However, for several cases, the GA is operated with 20 or 30 generations to ensure that the GA algorithm would not lead to further improvement of the optimum fitness function.

For our PSO method (Bonyadi et. al., 2017), the optimization of 2PA and 3PA signals is implemented with a swarm size of 1000 and a 9-dimension search space (the number of variables per particle to optimize – equivalent to the number of genes in a GA), the maximal number of generation is set to 40 or 60 (depending on the case) to ensure finding the optimal value of the fitness function, after which the algorithm is stopped (when it is not automatically interrupted beforehand when no further improvement of the optimum fitness function is observed).

## 3 Results

### 3.1 Selective wavelength optimization with single pulse seed

The spectral broadening obtained via single femtosecond pulse propagation is induced by well-known soliton evolution dynamics in an anomalous dispersion regime. Such an evolution typically yields the following scenario: solitons are formed at the onset of fiber propagation via conjoint yet opposite dispersive and nonlinear effects. The perturbations from ideal (high-order) soliton dynamics - seeded by high-order dispersion, Raman scattering, attenuation, etc. - lead to soliton fission, so that the initial pulse breaks-up to form and eject multiple individual solitons. During further propagation, the ejected solitons are shifted toward longer wavelengths via soliton self-frequency shift (SSFS) - a result of Raman scattering. Correspondingly, the spectral components at shorter wavelengths, located in the normal dispersion regime, are generated by dispersive waves directly radiated at the trailing edge of the solitons. For example, **Figure 3** presents the spectro-temporal evolution of a pulse with 12 kW peak power, 100 fs pulse duration, and chirp factor $C = -5$. The soliton fission occurs after around 30 cm of propagation, thus resulting in sudden spectral broadening. Soliton shifting contributes to further expanding the spectrum toward the longer wavelength until SSFS saturation (Dudley et al., 2006). Dispersive waves are generated at wavelengths around 1100 nm - 1200 nm. Any variation of the initial parameters (i.e., peak power, pulse duration, and chirp factor) brings about a noticeable alteration in the spectral broadening process, resulting in a change of MPA signals. However, SC generation induced by a single pulse seed provides only single-pulse-like MPA signals, as illustrated in **Figure 3C**.

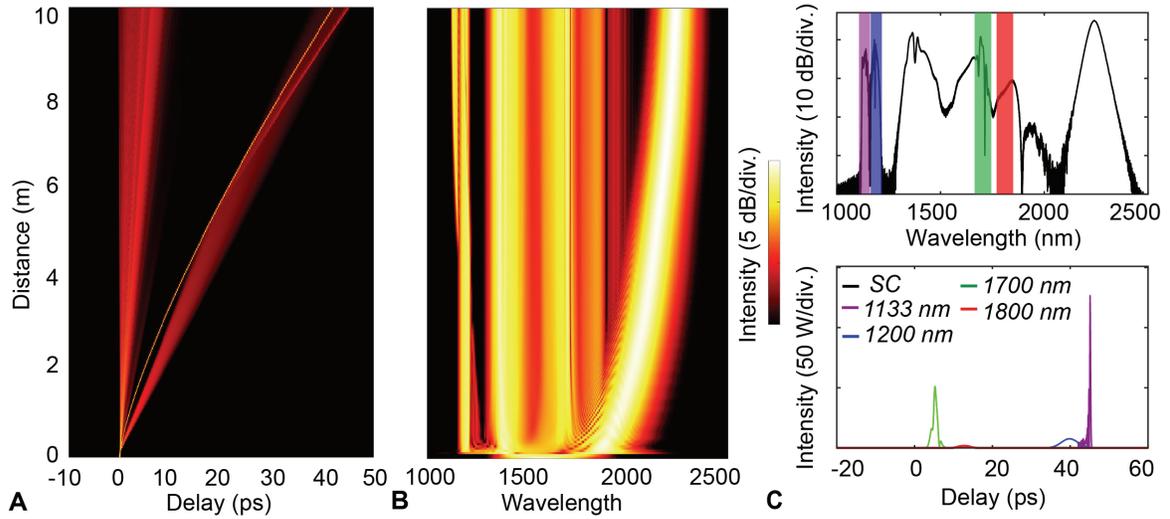

**FIGURE 3** | Numerical simulation of SC generation in 10 m of HNLF using a single pulse ($P_0 = 12$ kW ; $T_0 = 100$ fs ; $C = -5$). **(A)** Temporal evolution, **(B)** spectral evolution, **(C)** output spectrum, and corresponding temporal waveforms filtered at selected wavelengths.

In order to assess the tunability of these MPA signals generated from SC generation, we have performed extensive simulations with adjustable input parameters. **Figure 4** presents the evolution of maximal MPA signals (over successive sets of 1000 simulations) obtained by single pulse seeds with varying initial parameters (i.e., peak power, pulse duration, chirp factor).

We here compare numerical results obtained from GA optimization and Monte-Carlo method: Monte-Carlo simulations are implemented with $10^5$ random sets of initial parameters and we observe a progressive increase of MPA signals as we probe further the initial parameter space. GA optimization, on the other hand, brings on remarkable benefits: higher MPA signals can be readily obtained and, more importantly, for a given number of simulations, the MPA signals obtained from GA optimization are always better than random Monte-Carlo ones (making our GA approach computationally efficient, with a smaller number of simulations required to reach a desired MPA signal level). Of course, both the efficiency and convergence speed of GA optimization depend on the selected target wavelengths, and thus of the nonlinear dynamics involved. For instance, 3PA signals excited by 1800 nm and enacted by soliton shifting are maximized with 4000 simulations (see **Figure 4D**), while 2PA signals generated by 1133 nm and induced by dispersive wave are rather maximized with 14000 simulations (see **Figure 4A**).

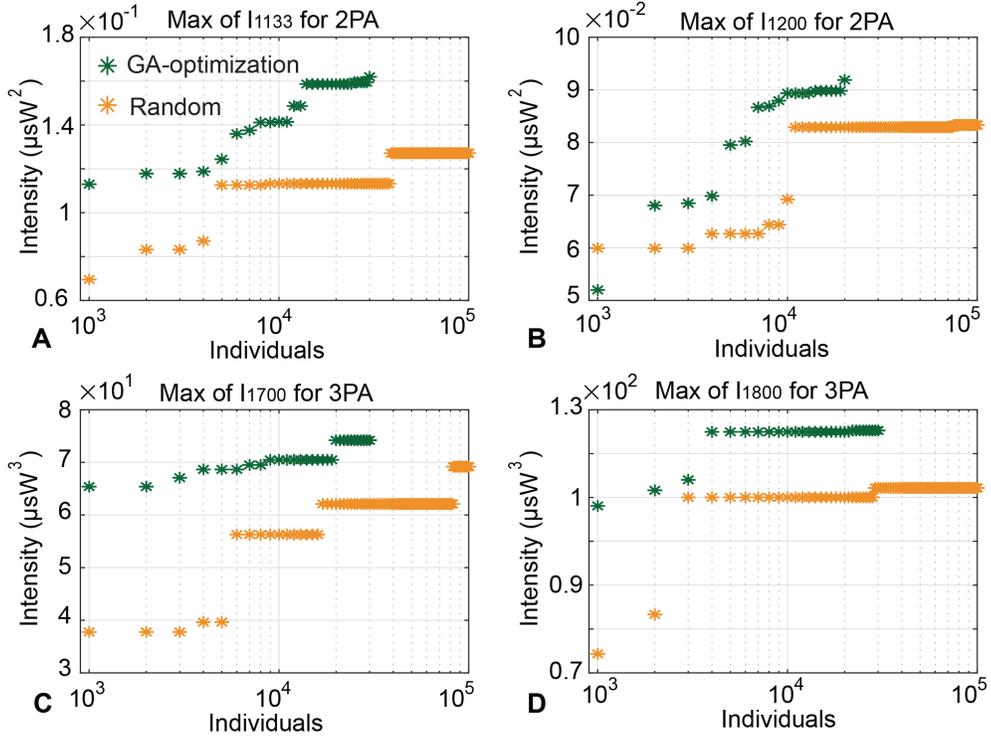

**FIGURE 4** | Evolution of maximal intensity for MPA signals with single pulse seeds: **(A)** 2PA excited at 1133 nm, **(B)** 2PA excited at 1200 nm, **(C)** 3PA excited at 1700 nm, and **(D)** 3PA excited at 1800 nm.

### 3.2 Selective wavelength optimization with reconfigurable input pulse patterns

The spectral broadening obtained from multiple femtosecond pulses propagation is, comparatively, more complex than "standard" soliton evolution dynamics associated with single pulse evolution. Indeed, each individual pulse within a reconfigurable pulse pattern (as generated from the approach illustrated in **Figure 5**) possesses a different peak power and duration. In fact, depending on the splitting ratio of the on-chip PDL, respective pulses can exhibit different nonlinear dynamics for spectral broadening, yielding various interactions over propagation and, ultimately, leading to different contributions in MPA signal generation (in the pure sense of complexity).

At the beginning of propagation, all individual pulses are spectrally broadened by self-phase modulation (SPM) and several ones, with sufficiently large peak powers (> 62 W), will generate solitons due to the effects of dispersion and nonlinearity. Within them, pulses with higher peak powers (> 245 W) can create a high-order soliton, which will subsequently break up into individual solitons via soliton fission. The ejected solitons have different durations, depending on the peak power of the individuals, and they respectively experience different self-frequency shift (i.e., solitons with higher powers would shift toward longer wavelengths), see **Figure 5**.

In such a scenario, complex inter-pulse interactions are expected to occur during further propagation: the temporal overlap between solitons ejected from different pulses, experiencing frequency shift at different rates, may occur at various fiber propagation distance. This can in turn lead to soliton collisions with energy exchange between the colliding solitons yielding the formation of new frequency components due to collision-induced dispersive waves (Erkintalo et.al., 2010; Luan et al., 2006).

Noteworthy, even the contribution of low power input pulses within the initial temporal pattern (i.e. not yielding soliton formation) or the lower intensity components generated during propagation (i.e. radiated dispersive waves, etc.) may also play a key role in the nonlinear mixing dynamics, especially when there are overtaken by other faster/slower frequency components radiated from adjacent pulses.

In this framework, such collisions dynamics can in principle be finely tuned by coherent control of the initial pulse patterns, so that to influence the power, duration, soliton frequency shift rate and relative phase between adjacent pulses, and thus adjusting the wavelength, propagation distance and timing of complex nonlinear conversion processes.

From a practical viewpoint, the adjustment of these multiple interactions may prove difficult. However, the suitable control of the initial parameters of the reconfigurable pulse patterns can provide a great variety and tunability to the excitation waveforms used for MPA.

For instance, a pulse pattern can provide MPA signals at all four selected wavelengths, yet in this case all excitation pulses filtered out from the SC essentially behave as single-like pulses with low peak powers, **Figure 5A1-5A3**. Other pulse patterns can be used for selective yet conjoint MPA processes: joint 2PA-3PA processes for a single fluorophore with an excitation wavelength at 1200 nm and 1800 nm can be achieved (**Figure 5B1-5B3**). Conversely, the excitation of dual 3PA processes (at 1700 nm and 1800 nm respectively) can be obtained so that to trigger a hybrid fluorescent medium including two distinct fluorophores (**Figure 5C1-5C3**).

The use of reconfigurable patterns brings on a flexible way to control the pulse shapes (i.e., temporal waveforms of the filtered pulses) toward a versatile reconfiguration and specific performance of the excitation pulse to be used for MPE microscopy.

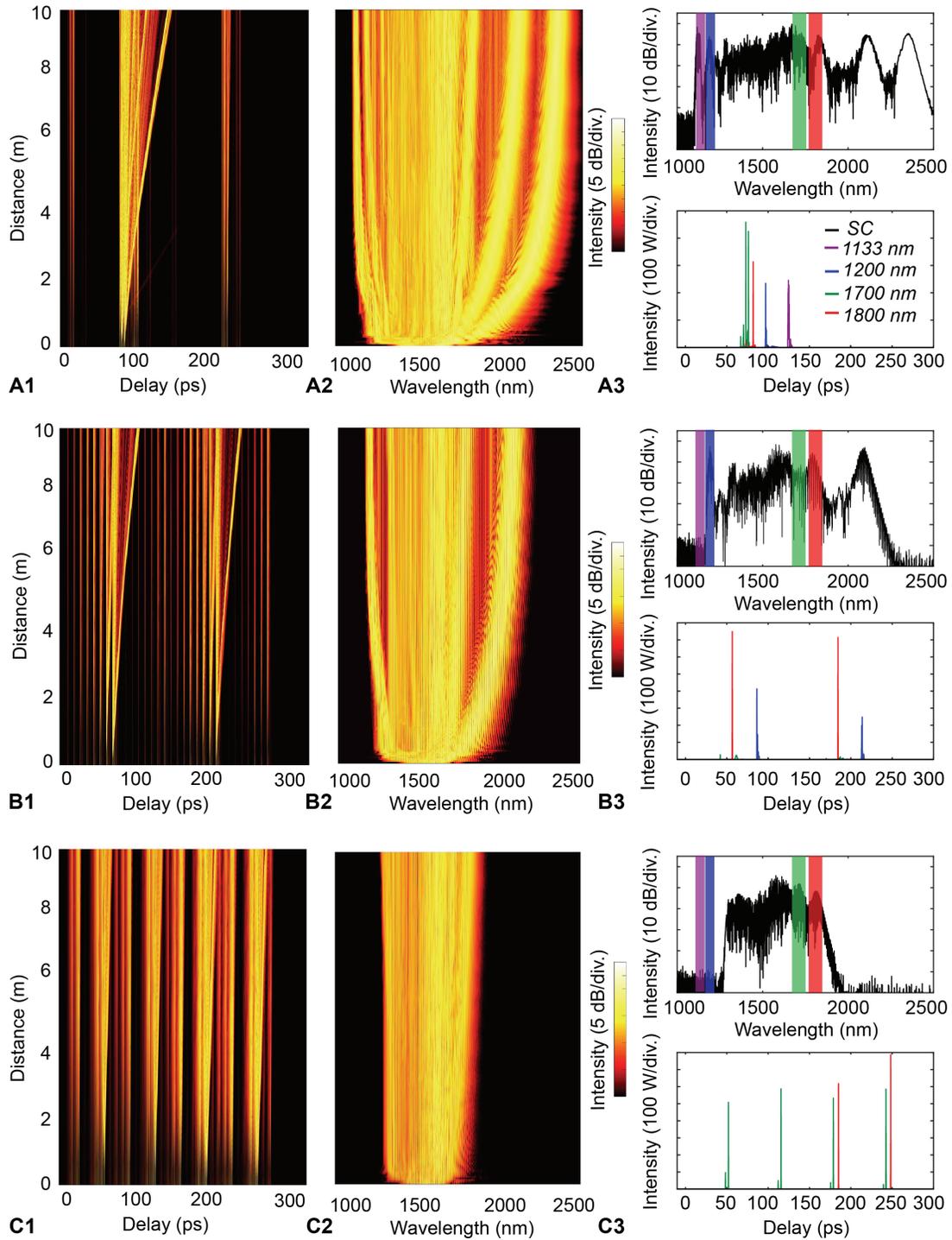

**FIGURE 5** | SC generation in 10 m of HNLF with various reconfigurable input pulse patterns for the excitation of (A1-A3) MPA signals at all four selected wavelengths, (B1-B3) conjoint 2PA-3PA of a single fluorophore with excitation wavelengths at 1200 nm and 1800 nm, (C1-C3) conjoint 3PA-3PA of different flurorophores with excitation at 1700 nm and 1800 nm.

As shown in **Figure 6**, while single input pulse seeds yield only single-like pulse waveforms for the desired wavelengths at the fiber output, the use of reconfigurable pulse patterns can offer both single-like pulse and more complex output waveforms with multiple pulse bunching operations. In fact, the delay and peak power of the individual pulses of the excitation waveforms can be effectively tailored by the modification of the input pulse patterns (i.e., only by changing the splitting ratio of the PDL). This brings about a potential to widen the modalities and selectivity for MPE and fluorescence lifetime microscopy.

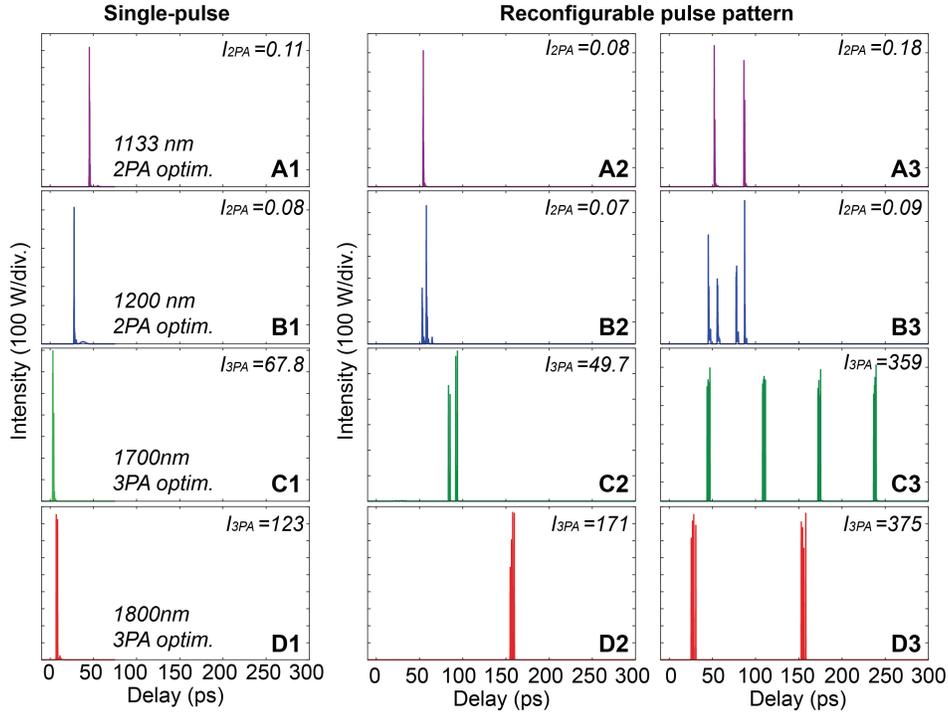

**FIGURE 6** | Selected temporal waveforms of filtered pulses with single pulse seed and the reconfigurable pulse pattern at selected wavelengths: waveforms filtered at 1133 nm (A1-A3), at 1200 nm (B1-B3), at 1700 nm (C1-C3), and at 1800 nm (D1-D3). $I_{2PA}$ and $I_{3PA}$ signals are here given in $\mu sW^2$ and $\mu sW^3$, respectively.

In **Figure 6**, we present examples of filtered output waveforms selected from Monte-Carlo simulations using random initials conditions (see section 2.1). These post-selected results clearly illustrate the variety and reconfigurability brought about by using tailored input pulse patterns instead of single pulse with adjustable properties. However, we note that the intensities of the excitation pulses at the selected wavelengths can be readily maximized using different algorithmic approaches such as GA or PSO. These optimization algorithms enable obtaining high excitation intensities with a limited number of simulations, which is the approach we explore in the following sections. We also note that a comparison between these optimization techniques with advantages and limitations is further discussed in section 3.4.1.

### 3.3 Selective multiphoton process and fluorophore optimization

The flexibility in terms of SC properties optimization, offered by reconfigurable input pulse patterns, allows in fact for customizing both the strength and trade-off between different MPA processes. For instance, a single broadband waveform can be used for the conjoint excitation of various MPA signals. This can be implemented for enhancing conjointly both 2PA and 3PA processes for a selected fluorophore, or conversely for the enhancement of dual 2PA-2PA or 3PA-3PA processes towards the combined excitation of two different fluorophores.

The results of such optimizations are presented in **Figure 7**, where we show the distribution of target MPA excitation intensities at selected wavelengths for various input pulse patterns. We compare the target function intensities (see section 2.2) obtained by using either a multi-objective GA optimization (to find a corresponding Pareto front) or random simulations using a Monte-Carlo method.

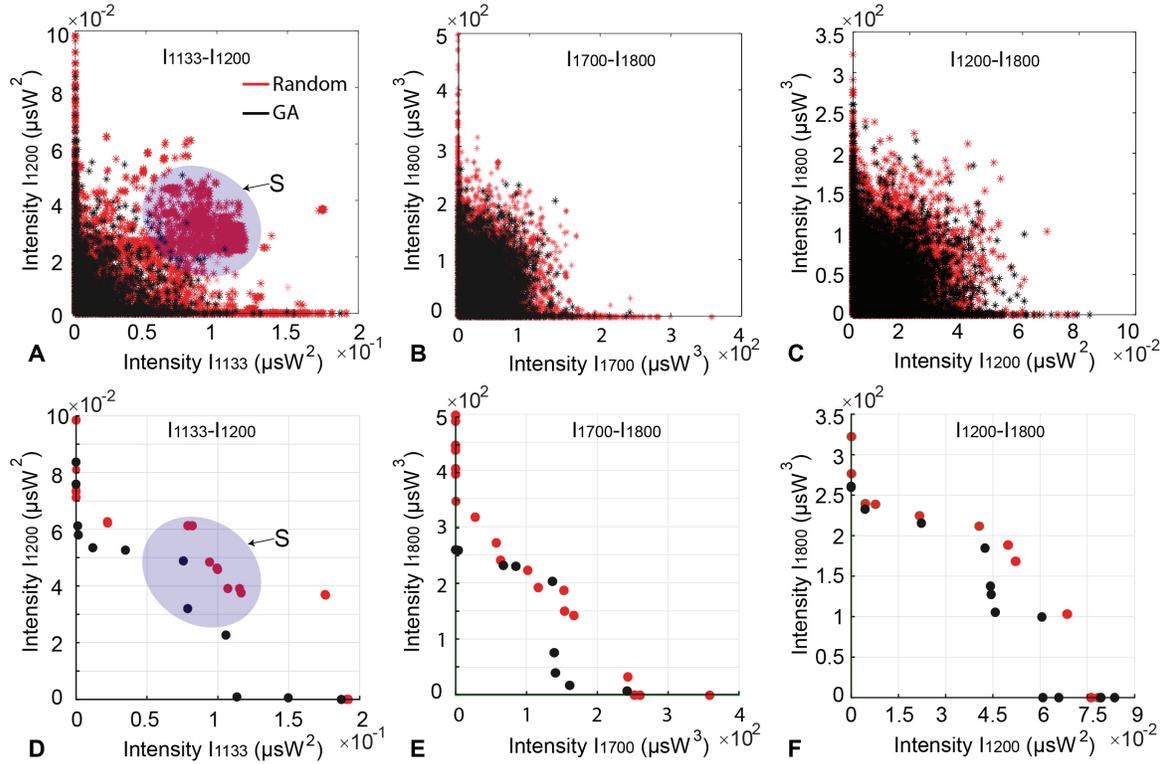

**FIGURE 7** | Distributions and Pareto fronts of MPA excitation intensities obtained from multi-objective GA and random simulations at selected target wavelengths: (A) and (D) at 1133 nm – 1200 nm, (B) and (E) at 1700 nm – 1800 nm, (C) and (F) at 1200nm – 1800 nm. The blue shaded area (labeled "S") illustrates a region of the Pareto front where the GA yields subtential improvements compared to Monte-Carlo simulations.

MPA excitation intensity distributions for different combinations of target wavelength are shown in **Figure 7A-7C**. From the shape of these distributions (i.e. clustering at the bottom left corner), one can see that the optimization of dual wavelength objectives consists in a trade-off between two optimal points of single wavelength MPA signals. In other words, as the system energy is bounded from the initial conditions with constant power, it appears (unsurprisingly) difficult to obtain conjoint MPA process signals at the same levels as the ones obtained from their single MPA process optimization.

This general rule here applies regardless of the optimization technique, however, GA optimization is useful to reach similarly high target MPA excitation signals with a smaller number of simulations. More importantly, GA optimization enables to enlarge the Pareto fronts compared to the Monte-Carlo method, **Figure 7D-7F**.

This aspect is important to tailor excitation waveforms and enhance the signals of conjoint MPA processes, which can be significant in several cases, as seen in the "*S*" shaded area in **Figure 7A**. Along with the potential to reconfigure individual excitation waveforms for distinct MPA processes, this approach is seen as an excellent way to achieve sophisticated characterization of various MPA processes and fluorophore combinations using the same experimental equipment and laser source.

To attest of this fine spectro-temporal waveform control, the filtered temporal profiles for conjoint MPA processes optimization are shown in **Figure 8** – here obtained for either single pulse or reconfigurable pulse patterns as input seeds.

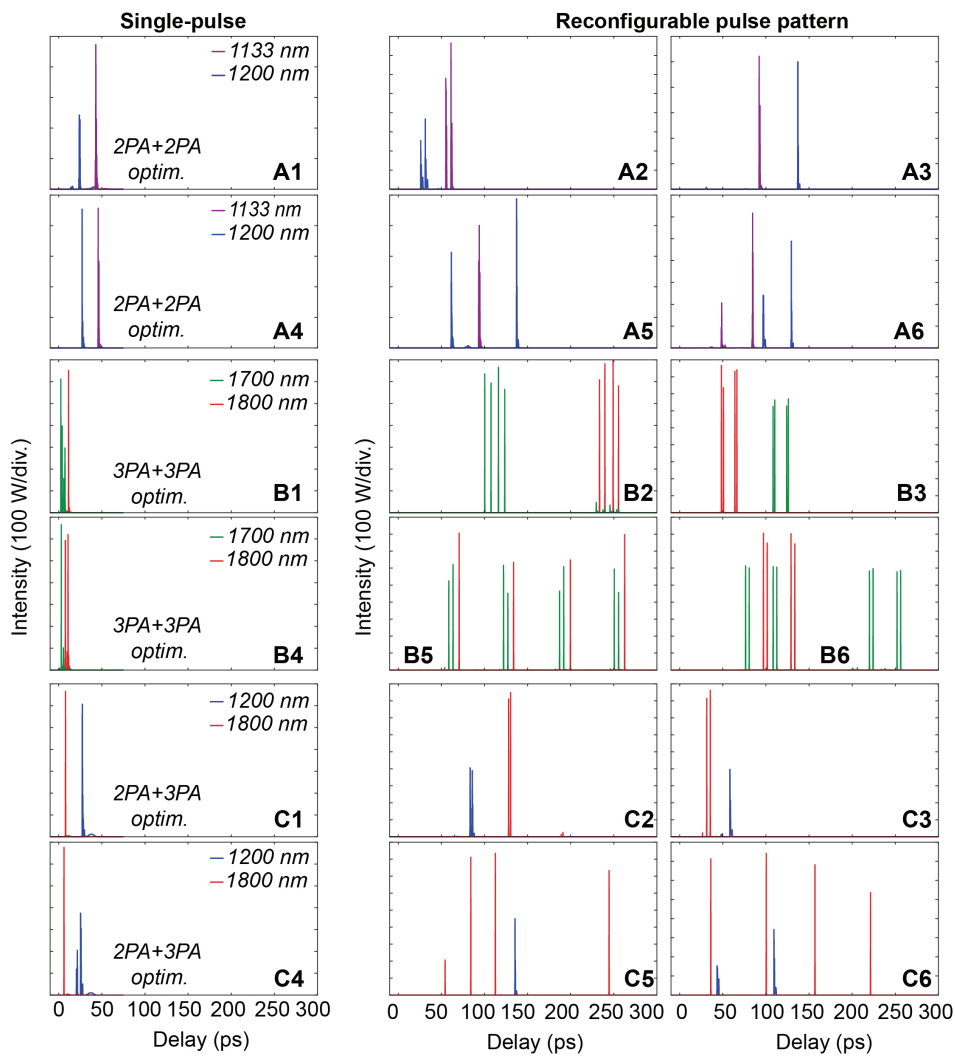

**FIGURE 8 |** Examples of temporal waveforms optimized by either single pulse seeds or reconfigurable pulse patterns for dual MPA processes enhancement at the selected wavelengths: (A1-A6) at 1133 nm and 1200 nm, (B1-B6) at 1700 nm and 1800 nm, (C1-C6) at 1200 nm and 1800 nm.

The use of single pulse seeds imposes a clear limitation for controlling the filtered MPA temporal profiles, typically associated with e.g. fixed relative delays between signals at selected wavelengths. In contrast, the MPA excitation waveforms are effortlessly tailored by using reconfigurable pulse patterns as input seeds. Using this approach, one can achieve the tunable adjustment of relative temporal delays between filtered signals, **Figure 8A2-A3, 8B2-B3, 8C2-C3**, but also enact more complex reconfiguration of the temporal waveforms to reach particular performances: selective pulse pattern adjustments can provide a variety of waveform excitations, for instance allowing for the formation of temporally-interleaved MPA signals, **Figure 8A5-A6, 8B5-B6, 8C5-C6**.

The ability of customizing the temporal waveforms for conjoint and/or selective MPA signals thus provides an innovative approach to develop and extend the potential for controlling complex multiphoton imaging techniques. Specifically, we expect that an improved tunability of the generated waveforms can prove useful to implement sequential spectral illuminations and time-encoded multiphoton processes using a versatile and scalable approach suitable with e.g. advanced STEAM or FLIM microscopy techniques. In a similar fashion, the use of reconfigurable output waveform bunches may mitigate several aspects associated to optical toxicity, by easily adjusting the equivalent peak power and repetition rate of the illumination (e.g. to alleviate either optical damage or thermal effects in the sample). Last but not least, tailoring the power spectral density and temporal waveform shape between different MPA processes may also be a way to indirectly adjust the penetration depth of a tissue sample. Indeed, it is known that 2PA processes for a selected fluorophore typically require less excitation power than ones of 3PA. However, with the use of a shorter excitation wavelength, 2PA yields higher scattering loss and thus lower penetration depth. Therefore, the idea of using reconfigurable pulse patterns for conjoint 2PA-3PA process optimization brings about a promising approach for "on-the-fly" imaging depth control (potentially without any moving parts).

### 3.4 Extension of optimization methods for multi-photon microscopy

### 3.4.1. Discussion on metaheuristic methods

We have seen that the use of reconfigurable pulse patterns gives rise to an extended parameter space so that to optimize a variety of MPA processes, similar to using an additional degree of freedom typical of multidimensional systems (here in the temporal domain, but analogous to e.g. the spatial dimension in multimode fibers). However, the efficiency of the targeted waveform intensity enhancement at selected MPA wavelengths depends strongly on both the algorithm used for optimization and the nonlinear dynamics involved during fiber propagation. **Figure 9** presents the evolution of optimal excitation intensities for 2PA and 3PA, considering an optimization based on either PSO, GA, or Monte-Carlo methods (using $10^5$ random simulations). In the fiber propagation regime considered (see section 2.1), the radiation of dispersive waves responsible for the generation of short wavelength components in a normal dispersion regime is less sensitive to the modification of the initial conditions than the generation of long wavelength components in the anomalous dispersion regime, here obtained via soliton shifting. In the former case (i.e. at 1133 nm and 1200 nm), metaheuristic optimization methods (GA or PSO) are not extremely significant compared to Monte-Carlo method to improve the target excitation intensity. In the latter case (i.e. at 1700 nm and 1800 nm), however, metaheuristic optimization techniques can prove very efficient to quickly reach optimal excitation intensities (at least twice higher than the ones from the Monte-Carlo method). For example, the maximal excitation intensity at 1800 nm obtained by PSO optimization is 630 μsW$^3$ after only 60 000 simulations (i.e.

when the algorithm stops after convergence) while $10^5$ random simulations only provide a maximum intensity of 260 $\mu sW^3$.

The results in **Figure 9** also provide a comparison of optimization methods used to optimize MPA signals. In all the considered cases, GA reach an optimal value relatively quickly and within $10^4$ simulations (after which the GA optimization does not seem to provide further significant improvement). While these results may be finely tuned by the adjustment of the GA parameters (e.g. population size, crossover, etc.) to better explore the available parameter space tune, however, PSO can readily achieve better signal enhancement in all four tested cases, but at the expense of increased computing power and duration (i.e. larger set of simulations before reaching convergence).

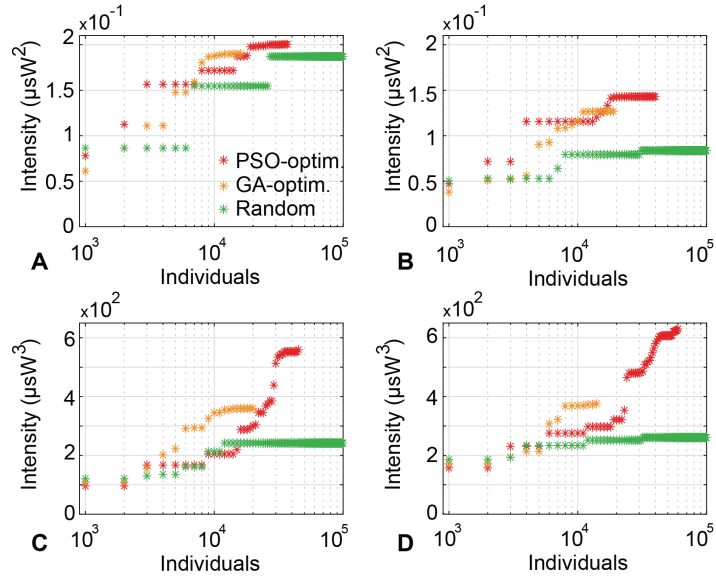

**FIGURE 9** | Evolution of maximal excitation intensity for MPA signals using reconfigurable pulse patterns, compared for different optimization approaches. (A) The intensity at 1133 nm wavelength for 2PA, (B) at 1200 nm for 2PA, (C) at 1700 nm wavelength for 3PA, (D) 1800 nm for 3PA.

The difference of efficiency between GA and PSO is caused by their principle (i.e., algorithm) to evaluate the best value for each iteration. GA relies on genetic evolution, and it starts with an initial parameter set of individuals (population). An iteration would be implemented if the termination criteria (e.g., number of generations) are not met so that the parameters of new individuals (i.e., children) are determined by interchanging gene groups of two selected current individuals (i.e., parents) through a crossover fraction (0.5 for our model). GA is usually a faster algorithm to find an optimum than other optimization algorithms, however, it does not guarantee to reach the global optimum of the system and may be stuck within a local minimum problem when exploring the parameter space.

PSO is, in contrast, modelled by the behavior of an animal swarm, starting with an initial swarm size of particles in $N$-dimensional search space (here, $N = 9$). The position in the search space of the next particles (i.e., the particle's trajectory) is updated by *swarm intelligence* that is determined by the current particle position, its individual-best position and the overall-best position of the particles in its neighborhood, following an algorithm regulated by self- and social-adjustment weights (Kennedy and Eberhart 1995). As GA, PSO does not guarantee to converge towards the global optimum, an issue that can also be mitigated by the adjustment of the PSO parameters (e.g. larger swarm size, values of self- and social-adjustment weight, etc.).

However, while GA determines the next individuals only from the selected current individuals (with marginal mutation that can be further implemented), PSO creates the next particles position not only from current particles but also from individual-best and overall-best positions.

In other words, PSO essentially operates with higher degrees of freedom than GA, thus providing a greater diversity in the particle trajectories and exploration of the parameter space than GA. In particular, the momentum effects on the particle movements can provide a faster convergence when a particle is moving in a directional gradient. Such a behavior can be readily seen in our results, for instance through the exponential increase of the MPA intensities at 1700 nm over the simulation iterations from 14000 to 30000, **Figure 9C**, and from 20000 to 40000 for MPA intensities at 1800 nm, **Figure 9D**.

### 3.4.2. Coherence optimization for multi-photon microscopy

The coherence degree of the excitation pulses (i.e. the shot-to-shot stability in both phase and amplitude) is a significant factor when considering multiphoton imaging techniques, which potential impact strongly depends on the practical applications targeted. For example, stability on a short time scale is not necessary for general purpose fluorescence microscopy applications, since the acquisition time is much longer than the shot-to-shot fluctuation of the output spectra, and the results thus averaged over long timescales. In contrast, other applications, such as CARS, fast MPE microscopy (with high temporal resolution), as well as most pump-probe measurements involving different wavelengths or frequency conversion phenomana require excitation pulses with high stability, and may even need excellent coherence if an intrinsically phase-dependent process is involved.

It is well-known that the coherence of SCs generated in an anomalous dispersion regime are intrinsically linked to the peak power and duration of input pulses (Dudley et al., 2006). If a single input pulse is used, the degree of control on the system overall coherence is thus strongly limited. On the other hand, the use of a reconfigurable pulse pattern enables the adjustement of the coherence by modifying both the peak power and the potential interactions between each individual pulse during fiber propagation.

In such a case, we can therefore maximize the excitation intensities for 2PA and 3PA while optimizing the coherence of the spectral components filtered for MPA. These results are illustrated in **Figure 10**, for which we run GA multi-objective optimizations trying to enhance MPA signals while either minimizing or maximizing the coherence degree of the corresponding filtered components.

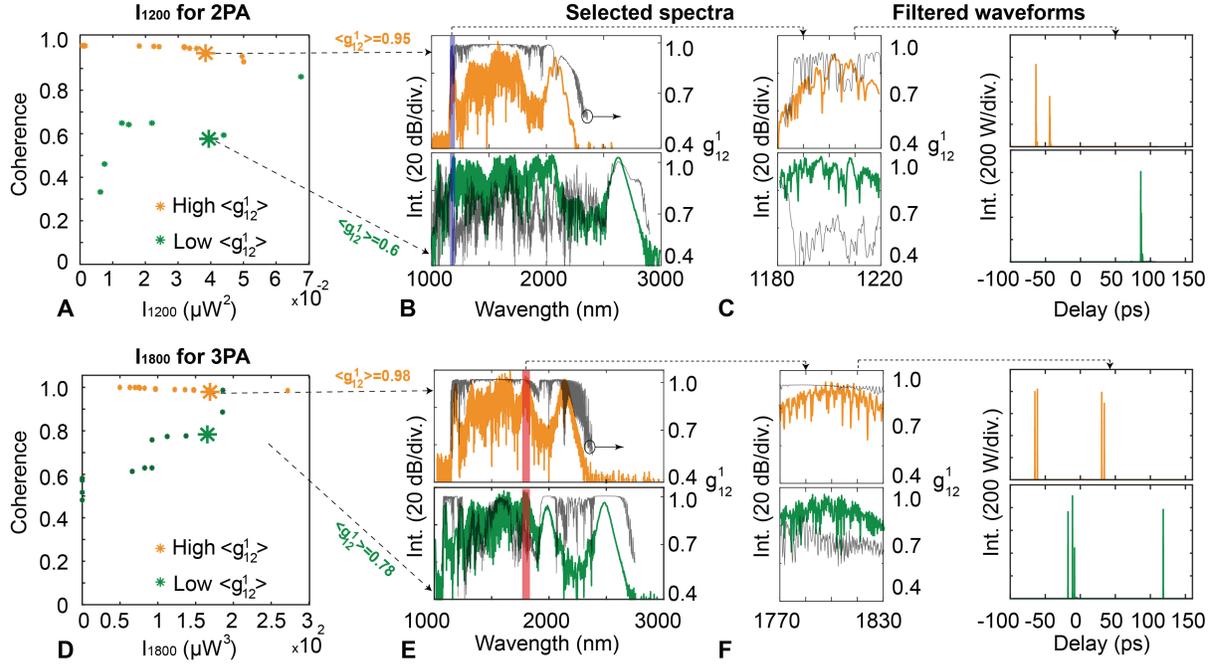

**FIGURE 10 |** (A) Pareto front of joint coherence and intensity optimization at 1200 nm for 2PA, (B) Selected SC spectra with a high and low coherence for 2PA, (C) Corresponding spectra and filtered waveforms at 1200 nm, (D) Pareto front of joint coherence and intensity optimization at 1800 nm for 3PA, (E) Selected SC spectra with a high and low coherence for 3PA, (F) Corresponding spectra and filtered waveforms at 1800 nm. The degree of coherence are shown in solid black line along the SC spectra of B, C , E and F (with scale on the right of the plots).

From the Pareto fronts of **Figure 10A** and **10D**, respectively for 2PA and 3PA optimization, one can see that an excellent coherence degree close to unity (with $<|g_{12}^1|>$ above 0.95) can be obtained in both optimizations with optimal MPA signals, while a lower value of coherence (with $<|g_{12}^1|> \sim 0.6$) can also be achieved with less than a two-fold reduction of the corresponding MPA signal levels. These optimized waveforms, presented in **Figure 10C** and **10F**, show that reconfigurable pulse patterns can indeed provide MPA excitation pulses with a similar intensity, but noticeably different coherence degrees and waveform shapes, opening further opportunity for accessing on-demand imaging modalities with versatile properties.

### 3.4.3. All-normal dispersion supercontinuum generation for multi-photon microscopy

For completeness, we also conducted similar optimization considering all-normal dispersion (ANDi) fiber propagation, a regime where spectral broadening is well-known for its inherent stability (Dudley et al., 2006; Sylvestre et al., 2021).

In our case, ANDi supercontinuum generation is simulated using Eq (1) with the same input reconfigurable pulse patterns and parameters as those provided in section 2.1 (i.e., pulse duration, peak power, amplification). For propagation, we consider a 2 meter-long homemade fiber that has typical near-zero, flat, all-normal dispersion in both the visible and near-IR (Huang et al., 2018; Canh et al., 2020; Le et al., 2021). At 1560 nm, our fiber has a nonlinear coefficient $\gamma$ = 13 W$^{-1}$km$^{-1}$ and the dispersion coefficients are $\beta_2$ = 6.36×10$^{-27}$ ps$^2$/m, $\beta_3$ = 3.2×10$^{-41}$ ps$^3$/m, $\beta_4$ = 3×10$^{-55}$ ps$^4$/m. In this regime, ANDi SC generation typically yields narrower spectral coverage than soliton-induced SC, requiring high peak powers for significant spectral broadening. As a consequence, we do not consider 2PA signals at 1133 nm and 1200 nm (that are too weak) for our optimization. However, for 3PA signals, multiple pulse optimization is also relevant for ANDi SC.

As shown in **Figure 11**, the maximum values of excitation intensities at 1700 nm and 1800 nm for 3PA are enhanced with GA optimization to reach 81 (µW$^3$) and 34 (µW$^3$), respectively. While both signals can eventually be enhanced conjointly, as seen from the Pareto front in **Figure 11B**, we note that the MPA signals are much lower than the ones obtained through soliton-based dynamics presented in section 3.4.1.

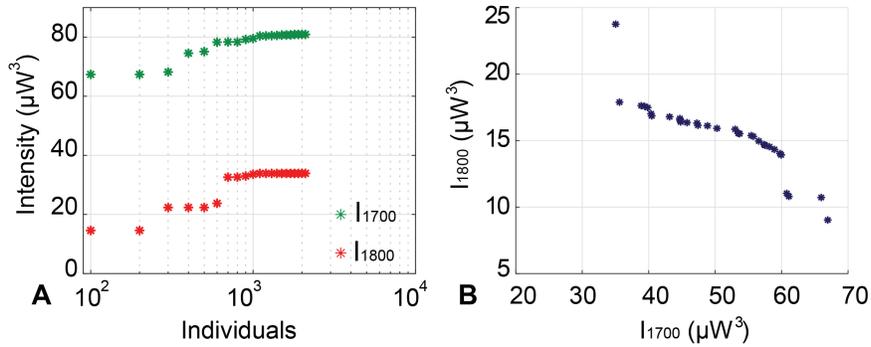

**FIGURE 11 |** The excitation intensity filtered from ANDi SC for MPA. (A) Evolution of maximum excitation intensity at 1700 nm and 1800 nm for 3PA, (B) Pareto front of optimized intensities at 1700 nm and 1800 nm for conjoint 3PA processes.

Importantly, ANDi SC generation is mainly driven by SPM broadening at the onset of propagation, and followed by optical wave breaking (OWB). Consequently, ANDi SC typically consists of a flat-top output spectrum paired with a uniform pulse shape (in the time-domain) with a relatively simple temporal profile. In the normal dispersion regime, modulation instability (the predominant phenomenon responsible for amplifying the vacuum noise during spectral broadening) is suppressed, resulting in output SC with significantly higher coherence (i.e., high shot-to-shot stability) than soliton-induced SC. However, ANDi regime here provides fewer degrees of freedom for efficiently controlling the spectro-temporal properties of SC generation, since OWB essentially occurs when there is a temporal overlap between SPM spectral components and the pulse tail (Heidt et al. 2011), so that during further propagation, no new spectral components are created. Therefore, the use of ANDi SC to maximize the excitation intensity for the targeted MPA processes is significantly limited.

## 4     Conclusion

Our results show a promising and feasible approach for controlling the spectro-temporal properties of excitation lasers for MPE microscopy: broadband SC generation covering numerous fluorophores absorption peaks (i.e. 2PA, 3PA) can provide a number of illumination wavelengths to match the proper excitation of selected fluorophores. By leveraging machine learning (GA, PSO), we can efficiently probe the input parameter space to selectively maximize particular output wavelengths, and thus rapidly enhance the desired MPA signals.

In particular, we show numerically that shaping an input wavepacket by means of an adjustable on-chip MZIs can be used to optimize a variety of temporal waveforms spectrally filtered from the output SC after propagation and spectral broadening. In fact, the on-chip MZIs can be used to create input pulse patterns with different properties (i.e., duration and peak power of the individual pulses) allowing for flexible tuning of the SC spectro-temporal properties for MPE microscopy. In contrast to single pulse seeds, reconfigurable pulse patterns provide a scalable and versatile way to adjust the output signals and temporal profiles at wavelengths directly useful for MPA processes (e.g. single-like pulses with high peak power, complex temporal profiles with multiple pulse bunching operations, etc.). From this approach, we demonstrate that the use of pulse patterns allows for controlling the delay and power between conjoint and interleaved MPA signals (e.g. conjoint 2PA-3PA processes for a single fluorophore, or conjoint 2PA or 3PA processes in a "mixture" of fluorophores).

Based on these promising numerical results, we expect that the practical implementation of these techniques will lead to further developments in MPE spectroscopy and multi-photon imaging techniques. With the flexibility to optimize excitation waveforms (in combination with machine learning), we anticipate that our approach will pave a way to enrich both the performances and modalities of MPE microscopy. While not exhaustive, such tailorable waveforms are seen as a possible way towards real-time control of MPE imaging depth, for conjoint multi-photon imaging techniques or tunable temporal interleaving of MPA signals (so that to observe living tissues and biological processes with minimal optical toxicity) or for direct image optimization and online analysis. It is worth noting that the dynamic illumination adjustment and combination of various degenerate and non-degenerate MPA processes holds a potential to probe spectral windows difficult to access from single wavelength excitation (e.g. low penetration depths from high scattering or water absorption regions) or to enhance the selectivity of particular modalities or fluorophore signatures towards adaptative image 'segmentation' of the sample content.

**Author Contributions**

VTH designed the study with contributions of BW. VTH performed the core numerical simulations, the data analysis and wrote the manuscript. YB and LS contributed to numerical simulations and data analysis. SF and VC provided input and contributed to discussion. BW supervised and coordinated the project. All authors contributed to the writing and revision of the manuscript.

**Acknowledgements**

This work has received funding from the European Research Council (ERC) under the European Union's Horizon 2020 research and innovation programme under grant agreement No. 950618 (STREAMLINE project). B.W. further acknowledges the support of the French ANR through the OPTIMAL project (ANR-20-CE30-0004) and the Région Nouvelle-Aquitaine (SCIR & SPINAL projects).